\documentclass[prb,twocolumn,showpacs,preprintnumbers,amsmath,amssymb]{revtex4}  
   
\usepackage{graphicx}
\usepackage{dcolumn}
   
\newcommand{\e}{\epsilon}
 
\newcommand{\D}{\Delta}
\newcommand{\s}{\sigma} 
\newcommand{\beq}{\begin{equation}}   
\newcommand{\eeq}{\end{equation}}   
\newcommand{\beqa}{\begin{eqnarray}}   
\newcommand{\eeqa}{\end{eqnarray}}

\newcommand{\br}{{\bf r}}   
   
\newcommand{\percent}{\%}   
\begin{document}   
\title{Energetic balance of the superconducting transition across the BCS-Bose Einstein crossover 
in the attractive Hubbard model}  
\author{A. Toschi$^{1,2}$, M. Capone$^{1,3}$, and C. Castellani$^1$}    
\affiliation{$^1$ Dipartimento di Fisica, Universit\`a di Roma ``La Sapienza'', and  
INFM-SMC, Piazzale Aldo Moro 2, I-00185 Roma, Italy}  
\affiliation{$^2$ Max Planck Institut f\"ur Festk\"orperforschung, Heisenbergstr. 1, 70569, Stuttgart, Germany}
\affiliation{$^3$ Istituto dei Sistemi Complessi del CNR, Via dei Taurini 19, I-00185, Roma, Italy}
  
\begin{abstract}  
We investigate by means of Dynamical Mean-Field Theory 
the crossover from BCS superconductivity to Bose-Einstein (BE) condensation of
preformed pairs which occurs in the attractive Hubbard model by increasing the attraction
strength.
We follow the evolution of the two energy scales underlying 
the superconducting phenomenon, the gap $\Delta_0$ and the superfluid stiffness $D_S$, which 
controls the phase coherence.  The BCS-BE crossover is clearly  mirrored in a change in the 
hierarchy of these two scales, the smallest of the two controlling the critical temperature. 
In the whole intermediate-to-strong coupling region $T_c$ scales
with $D_S$, while $T_C$ is proportional to $\Delta_0$ only in the BCS regime.
This evolution as a function of the interaction qualitatively resembles what happens in the cuprates
when the doping is decreased towards the Mott insulator.

This continuous change reflects also in the energetic balance at the 
superconducting  transition. While, as it is well known, superconductivity 
is stabilized by a potential energy gain in the BCS regime, the strong-coupling 
superconductivity is made stable by a reduction of kinetic energy. 
Interestingly the intermediate-coupling  region, 
where the maximum $T_c$ is achieved, behaves similarly to the strong-coupling
regime, and its gain in kinetic energy is the largest as a function of the coupling.
Since the integral of the optical conductivity is proportional to the kinetic energy,
the above finding implies that the attractive Hubbard model can account qualitatively for the 
anomalous behavior of optical spectra around $T_c$, where an increase of spectral weight
is observed in under and optimally doped cuprates, while the overdoped samples have a more
standard behavior.
This qualitative agreement is lost in the normal phase, specifically at strong-coupling, calling for
the  inclusion  of strong correlation effects in the theoretical description. 
\end{abstract}  
  
\date{\today}   
\maketitle   
  
\section{Introduction}  
High-temperature superconductors (HTSC) have attracted an unprecedented interest 
in the solid-state community, yet the number of open issues  largely exceeds the 
number of generally  accepted points. Among the latter, we can count the crucial 
role of strong electron-electron repulsion. This immediately raises the question 
of the competition and coexistence between this repulsion and the 
 attraction between quasiparticles responsible of the superconducting phenomenon. 
 
In this work, we make a step back, and we consider a simpler and more  
``phenomenological'' question, that is the possibility to explain some  
features of the HTSC physics simply in term of a crossover between a  
conventional (BCS) superconductivity (characterizing the overdoped compounds)  
and a sort of strong-coupling Bose-Einstein (BE) superconductivity  
(in the underdoped region).\cite{micnasrev,varie}  
A number of properties of the cuprates can in fact be qualitatively 
described in this framework. Notable examples are the
intermediate values ($\sim 10\div 20$ \AA) of the  superconducting coherence length  
$\xi_0$ in optimally doped compounds\cite{pan,iguchi}, and 
the pseudogap phenomenology in the  underdoped compounds\cite{timusk},
which can be ascribed to a ``preformed pairs'' phase, in which the
superconducting order parameter has a finite amplitude, but it 
lacks  phase coherence.
The strongest connection with the physics of a BCS-BE crossover is perhaps the evolution with doping  
of the relevant energy scales controlling the stability of the superconducting
condensate,  
namely the gap $\Delta_0$, which is proportional to the binding energy  
of the Cooper pairs, and the superfluid stiffness $D_S$, which  
represents the energy cost for phase fluctuations. In the underdoped  
compound there is a clear experimental evidence\cite{uemura1,ding} that  
$D_S < \Delta_0$ and $T_c$ is proportional to the ``weak link of the  
chain'' $D_S$ according to the so-called Uemura plot\cite{uemura},  
whereas for higher doping a more conventional direct proportionality of
$T_c$ on $\Delta_0$ is recovered.   
  
However some obstacles arise trying  to push further this line of thought.
 For instance,   
the unconventional $d-$wave symmetry of the order parameter, with the   
related presence of quasiparticle excitations down to zero energy, 
is known to strongly affect the low-temperature thermodynamic properties.
In particular the nodal quasiparticles can dominate the low-temperature 
charge\cite{arun,lara} and thermal   
transport even in the strong-coupling regime, invalidating a purely 
bosonic description.

It is also intuitively hard to reconcile the BCS-BE crossover scenario with the 
relevance of strong correlation effects, which are naturally larger and
larger when the doping is reduced and the Mott insulator is approached. 
Then, in the BCS-BE scenario, the attraction should be stronger in the 
same region where the repulsion is maximum. This puzzling fact can be
understood by interpreting the BCS-BE physics as relevant only for
{\it low-energy} quasiparticles, while the high-energy physics has
to be dominated by Coulomb repulsion. An explicit realization of 
a similar scenario has been obtained in models with orbital degeneracy and
specific kind of interactions\cite{capone1,capone2}, where
the superconducting properties of the strongly renormalized quasiparticles 
share similarity  with those
of an effective attractive Hubbard model\cite{capone1}.

Keeping in mind those limitations, we find it important to understand if, and to what extent,
 a simple attractive picture for quasiparticles, modelized by an attractive Hubbard
model, is able to reproduce some of the properties of the cuprates.
We will find that while some of the properties of the superconducting phase
can be put in this framework, the normal state behavior necessarily requires
the inclusion of strong repulsive correlations.
The Hamiltonian of the attractive Hubbard model reads
\begin{eqnarray}  
\label{hubbard}  
{\cal H} = &-&t \sum_{<ij>\sigma} c_{i\sigma}^{\dagger} c_{j\sigma}   
-U\sum_{i}\left ( n_{i\uparrow}-{1\over 2}\right )  
\left ( n_{i\downarrow}-{1\over 2}\right ),  
\end{eqnarray}   
where $<ij>$ indicates that the first sum is restricted to 
nearest neighbors sites only,
 $c_{i\sigma}^{\dagger}$ ($c_{i\sigma}$) creates (destroys)   
an electron with spin $\sigma$ on the site $i$ and $n_{i\sigma} =   
c_{i\sigma}^{\dagger}c_{i\sigma}$ is the number operator;   
$t$ is the hopping amplitude and $U$ is the Hubbard on-site attraction. 
As it can easily seen, this model 
reproduces in the extreme weak- and strong-coupling limit the  
BCS and BE regime respectively, and allows to move in the whole crossover   
region, simply  by adjusting a single parameter, the ratio $U/t$.

From the theoretical point of view, 
a stringent test of the relevance of the crossover in the attractive
Hubbard model for the cuprates
has been mainly limited by the lack of reliable non perturbative
approaches able to follow the evolution from weak to strong coupling 
without a bias in some direction. 
In this work we use for this purpose the Dynamical Mean-Field Theory (DMFT)\cite{dmft},
a non perturbative approach which  neglects the spatial correlations beyond the mean field level,   
but fully retains the local quantum dynamics, and becomes exact in the
infinite coordination limit.
The local nature of the attraction and the non-perturbative nature of DMFT 
are expected to lead to reliable results and allow for a democratic treatment 
of the different regimes and of the whole BCS-BE crossover.

The attractive Hubbard model has been already investigated with DMFT in the 
past, but the attention was focused mainly on the normal phase or on the 
determination  of the critical temperature for the superconducting transition,
with the recent exception of Ref. \onlinecite{randeria}, where the 
BCS-BE crossover at $T=0$ is analyzed using iterated perturbation theory
as an impurity solver, but the focus is mainly on spectral properties.  
More precisely, by explicitly avoiding superconducting solutions,  
a phase transition has been found both at finite\cite{keller,paolo}  
and at zero temperature\cite{prl} between a metal and a ``paired'' phase,  
i.e., a collection of independent pairs without the superconducting phase  
coherence. This paired phase represent the  
'negative-$U$' counterpart of the paramagnetic Mott insulator found for the  
repulsive Hubbard model\cite{prl,keller}.  
  
With the present work we complete the DMFT analysis of this model 
reported in the Refs. \onlinecite{prl,keller,paolo}  with a careful  
investigation of the properties of the superconducting phase, 
certainly stable at  low temperatures. 
Although the onset of the superconductivity smoothes the abrupt changes 
observed in the low-temperature metastable normal phase, and the 
evolution of the superconducting phase as the coupling is increased
is a smooth crossover, the way in which the energetics changes in 
this process is extremely interesting.
We fully characterize the crossover from BCS to BE superconductivity, 
and we establish that the intermediate regime, where the maximum $T_c$ is 
obtained, shares the behavior of the strong-coupling, BE
superconductivity. In particular, this means that the ``optimal'' superconductor
is stabilized by a kinetic energy gain, as it is expected in the bosonic regime,
as opposed to the standard potential energy stabilization characteristic
of weak-coupling superconductivity.
 
The paper is organized as follows: Sec. II briefly describes the DMFT
for an s-wave superconductor, Sec. III is devoted to the evolution
of the superfluid stiffness and the gap as a function of the interaction,
Sec. IV presents the energetic balance at the superconducting transition,
and the relation with optical measurements. Sec. V is dedicated to 
concluding remarks.

\section{method}  
In this section we will briefly introduce DMFT and some aspects
related to the study  of superconducting solutions.
DMFT maps a quantum lattice model onto an impurity model whose
hybridization function (usually called Weiss field in analogy with
classical mean-field theories) is determined by means a self-consistent
equation. The latter equation contains the information about the 
 original lattice only through the non-interacting density of states.
In our case (\ref{hubbard}) is mapped onto  
an Anderson model with attractive coupling. In order to describe the
superconducting phase, the bath presents superconducting terms leading
to an anomalous Weiss field
\begin{eqnarray}  
\label{aim}  
{\cal{  H}}_{AM} &  = &  \sum_{l,\sigma} \,   
\left[ \epsilon_l \, c^{\dag}_{l\sigma} \, c_{l\sigma} + V_l\,  
( c^{\dag}_{l\sigma} d_{\sigma} + \mbox{h.c.}) \right. \nonumber \\ 
&  + & \left. \Delta_l \,  
(c_{l\downarrow}^{\dag} c_{l\uparrow}^{\dag} + \mbox{h.c.}) \right] 
+ {\cal{H}}_{loc} 
\end{eqnarray}  
where $ {\cal{H}}_{loc} = -U\left ( n_{0\uparrow}-\frac{1}{2}\right )  
\left ( n_{0\downarrow}-\frac{1}{2}\right ) - \mu n_0$ is the on-site term, and 
the chemical potential $\mu$ is adjusted to fix the particle
density on the impurity site (in our calculations we always fix  $n=0.75$
as a generic density out of half-filling,  as done in  Refs. \onlinecite{prl,paolo}). 
 
From the impurity model we compute the normal and anomalous Green's functions, 
$G(\tau)=-\langle T c_{\uparrow}(\tau)  c^{\dag}_{\uparrow}(0)\rangle$ and   
 $F(\tau)=- \langle T c_{\uparrow} (\tau) c_{\downarrow}(0)\rangle$, which 
are used to build the matrix $\hat{G}(i\omega_n)$ in Nambu-Gor'kov formalism.
Analogously one can define a matrix of Weiss fields whose
diagonal ${\cal G}^0(i\omega_n)^{-1}$ and off-diagonal 
${\cal F}^0(i\omega_n)^{-1}$ elements are related to the  
parameters appearing in (\ref{aim}) by 
\begin{eqnarray}
\label{gf0} 
 {\cal G}_0^{-1}(i\omega_n) &  = & i\omega_n +\mu + \sum_{l=1}^{n_s}  
\, V_l^2  
\, \frac{i\omega_n+\epsilon_l}{\omega_n^2+ \epsilon_l^2 +\Delta_l^2} \nonumber
 \\ 
 {\cal F}_0^{-1}(i\omega_n) &  = &  + \sum_{l=1}^{n_s} \, V_l^2 \,  
\frac{\Delta_l}{\omega_n^2+ \epsilon_l^2 +\Delta_l^2}.  
\end{eqnarray} 
The two above quantity also define the local self-energy matrix
\begin{equation}
\hat\Sigma(i\omega_n) = \hat{\cal G}_0^{-1}(i \omega_n) - \hat{G}^{-1}(i\omega_n).
\end{equation}
 
The self-consistency condition relates the Weiss field to the Green's function.
We work with an infinite-coordination  
Bethe lattice with semicircular density of states of half-bandwidth $D$,
(i.e., $N(\epsilon)=2/(\pi D^2) \, \sqrt{D^2-\epsilon^2}$), for which the
self-consistency reads
\begin{equation}  
\label{selfbethe}  
\hat{\cal G}_0^{-1}(i \omega_n)= i\omega_n \hat{\tau}^0 + \mu \hat{\tau}_3 
 -t^2  \hat{\tau}_3 \hat{G}(i\omega_n) \hat{\tau}_3     
\end{equation}  
 $\hat{\tau}_{i}$ being the Pauli matrices. 
It is straightforward to check that
 Eqs. (\ref{aim})-(\ref{selfbethe}) reduce automatically 
to their normal-state  counterparts as soon 
$\Delta_l  = 0$ for each $l$.  
  
The inclusion of local quantum fluctuations
rules out the possibility to analytically solve  Eqs. 
(\ref{aim})-(\ref{selfbethe}), and requires numerical (or approximate)
solutions of the Anderson impurity  model (\ref{aim}). 
Here we adopt Exact Diagonalization (ED) as the impurity solver.  Thus, 
we discretize the Anderson model, by truncating the sums  
in Eqs. (\ref{aim}) and (\ref{gf0}) to a small finite number of levels  
$N_s$. It has been shown that extremely small values of $N_s$ provide   
really good results for thermodynamic properties and reliable results  
for spectral functions.\cite{krauth}  
Here we use both the Lanczos algorithm (at zero temperature) and the 
finite temperature algorithm in its simplest version, which requires
the full spectrum of the Hamiltonian matrix.
To obtain the full spectrum of the Hamiltonian, needed to compute the finite temperature 
properties, we are forced to a rather small value of $n_s$, namely $5$, 
whereas in the zero temperature case the Lanczos algorithm allow us to deal
with larger systems, up to $n_s \sim 10$ also in the superconducting phase,
where the number of particles is not conserved, making the size of the
Hilbert space larger.

\section{Hierarchy of energy scales}  

As we anticipated in the introduction, a landmark of the  
HTSC phase-diagram is the crossing of the energy scales relevant
for superconductivity: the superconducting gap $\Delta_0$ and 
the superfluid stiffness $D_S$, which represent respectively
the energetic cost to form a Cooper pair and to break the phase coherence
of the superconducting phase. 

DMFT allows for a straightforward 
calculation of the order parameter $\Delta_0$ at zero temperature, 
defined through the local anomalous Green's function 
\begin{equation}
\Delta_0= -U \langle T c_{\uparrow} (0) c_{\downarrow}(0)\rangle=
-U / \beta  \sum_n F(i \omega_n).
\end{equation}
We note that the values for $\D_0$ obtained with this standard ``mean-field'' definition 
coincides almost exactly with the anomalous part of the self-energy 
$\Sigma_{12}(i\omega_n)$  at large  $\omega_n$. In our approach $\Sigma_{12}$ is 
frequency dependent, with a small reduction at small Matsubara frequencies with respect
to the large $\omega_n$ value. In most cases we find that the above reduction is quite
small and $\Sigma_{12} \simeq  \Delta_0$.

On the other hand, the superfluid stiffness $D_S$, 
is defined in terms of the static limit of the 
electromagnetic response function as
\begin{equation}
D_S= D_{dia} - \chi_{jj}({\bf q} \rightarrow 0, \Omega=0)
\label{ds}.
\end{equation}    
The diamagnetic term $D_{dia}$ is given by
$D_{dia} = - \langle E_{kin} \rangle = - 2/\beta  \sum_{\omega_n} \int d 
\epsilon \epsilon N(\epsilon) G(\epsilon,\omega_n)  $\cite{notadiam}).
The paramagnetic term, which measures the normal component 
$D_N$ is defined as the sum over all the 
directions\cite{notadim} of the transverse part of 
the paramagnetic kernel in the static limit 
(i.e., $\chi_{jj}({\bf q} \rightarrow 0, \Omega=0)
= \sum_{\alpha=x,y,z,\ldots}\chi_{jj}^{\alpha\alpha}({\bf q} \rightarrow 0,
 \Omega=0) $ being $\chi_{jj}^{\alpha\alpha}(\br,\tau)=
\langle T_\tau \vec{j}_\alpha(\br,\tau) j_\alpha(0,0) 
\rangle$). 

In principle the evaluation of  
$\chi_{jj}({\bf q} \rightarrow 0, \Omega=0)$ would require
the calculation of the corresponding four-field correlation function, but 
a remarkable simplification occurs in the infinite coordination 
limit where the DMFT is exact, since all the
vertex corrections to the e.m. kernel vanish\cite{dmft}.
The evaluation of $D_S$ requires therefore only 
the dressing of the non-interacting Green's function in the electromagnetic
 kernel with the
self-energy. As a results, the DMFT expression for $\chi_{jj}$  reads\cite{dmft,chatt}
\begin{eqnarray*}
\chi_{jj} & = & -\frac{2}{\beta} \sum_{n} \int \, d\epsilon \, N(\epsilon) \, 
V(\epsilon) \\
&\times & \left[ G(\epsilon,\omega_n) G^{*}(\epsilon, \omega_n) +
F(\epsilon,\omega_n)F(\epsilon,\omega_n) \right]
\label{chijj} 
\end{eqnarray*}         
where $V(\epsilon)=(4t^2-\epsilon^2)/3$ is the square current vertex for
the Bethe lattice\cite{chung,chatt,bluemer} and $G(\epsilon,\omega_n)$, 
$F(\epsilon, \omega_n)$ are the normal and anomalous lattice Green 
function respectively \cite{notadefds}.

\begin{figure}[h!bt]
\includegraphics[width=8cm,angle=0]{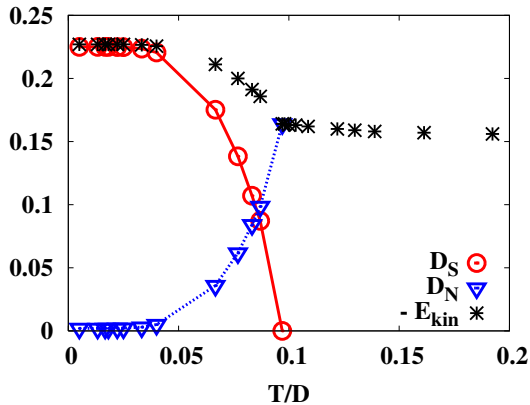}
\caption{\label{fig:sup_dens} 
(Color online) Temperature dependence of the superfluid stiffness  $D_S$ and of the 
normal component $D_N$ at $U=2.0 D$ (just before the maximum $T_c$)
 compared with the diamagnetic term $-\langle E_{kin} \rangle$.} 
\end{figure}

Thanks to the specific form of the current vertex, 
it is possible to write a more compact expression for  $D_S$. 
More specifically 
by exploiting the relation $ -\epsilon N(\epsilon) = \partial_\epsilon  [
V(\epsilon) N(\epsilon)]$ and then transferring the energy-derivative on the 
Green functions one finally gets
\begin{equation}
D_S = \frac{4}{\beta} \sum_{n} \int \, d\epsilon \, N(\epsilon) V(\epsilon) \;
F(\epsilon,\omega_n) F(\epsilon,\omega_n)
\end{equation}
The above expression only contains the anomalous Green functions, and makes 
it apparent the vanishing of $D_S$ at $T=T_c$. 

Before comparing the evolution of $\Delta_0$ and $D_S$, we briefly discuss
the temperature behavior of the superfluid and normal densities.
In particular, the evaluation of $D_N$ and $D_S$ 
reveals that the $T=0$ value of $D_N$ is
basically negligible, as predicted by BCS mean-field, not only in the weak-coupling
regime, but also for sizable interaction, as shown in Fig. \ref{fig:sup_dens} for
$U=2D$, where $D_N/D_S$ at $T=0$ turns out to be less than 0.01.
This means that it is possible to identify 
 $D_S(T=0)$ with $ - \langle E_{kin} \rangle $ in a wide range of couplings.
It is clear that this results depends on the neglect of small-momentum
collective modes, which can deplete the condensate\cite{lungo}.
It is worth noting that the presence of a frequency-dependent self-energy and of an 
incoherent part of the Green's function does not lead to a reduction to a depletion 
of  $D_S$, differently from what happens, for instance, for impurity scattering.

\begin{figure}[hbt]
\includegraphics[width=8cm,angle=0]{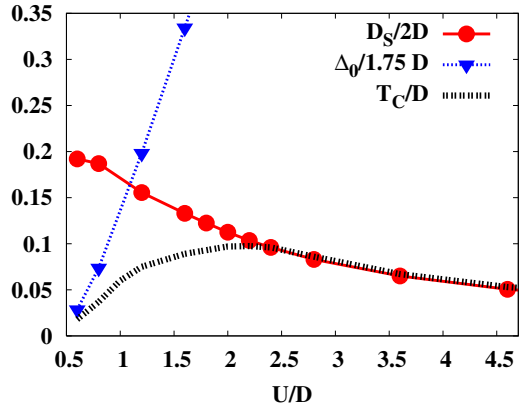}
\caption{\label{fig:cross_dmft} 
(Color online) Superconducting energy scales (in unit of $D$) ($\D_0$ and $D_S$ have been 
normalized to be directly comparable with $T_c$ in the asymptotic regimes).
The two energies cross  at intermediate coupling at $U \sim D$, 
well before the maximum $T_c$ dome ($T_c$ is taken from Ref. \onlinecite{paolo})} 
\end{figure}

We now come to the evolution of the energy scales inherent to the superconducting phase
as a function of the interaction.
Our results are summarized by  Fig. \ref{fig:cross_dmft}, where the  
superconducting gap $\Delta_0$, the superfluid stiffness $D_S$ and the critical temperature $T_c$
are plotted as a function of the ratio $U/D$. $\Delta_0$ is smallest in the weak-coupling regime,
and the superfluid stiffness becomes the lowest scale in strong-coupling.
It is natural that the system becomes superconductor only when the pairs are formed and 
they have phase coherence, namely when the temperature is lower than both the gap scale and 
the superfluid stiffness scale. Therefore the critical temperature is substantially determined
by the smaller scale at each interaction. 
Thus the  critical
temperature is proportional to $\Delta_0$ in the weak-coupling regime, as predicted by BCS mean-field,
while  in the strong-coupling limit we recover $T_c \simeq D^2/4U \simeq \langle  E_{kin}\rangle  /2  \sim D_S/2
$, as predicted by the  mean-field  solution in the limit of an hard core boson system
with Heisenberg coupling $J=D^2/(d U)$\cite{micnasrev,notadensity}

The crossing between the two energy scales, which is   characteristic of the BCS-BE crossover 
 occurs at intermediate coupling for  $U$ slightly smaller than maximum 
$T_c$ value.
 Therefore only in the proximity of the weak coupling regime the smallest energy scale, which ultimately determines $T_c$,
is the binding energy of the Cooper pairs, 
whereas already at the optimal $U$ (and of course in all the strong coupling regime) 
the weak link for the stability of the superconducting phase is  the superfluid stiffness
$D_S$ (hence, the onset of the superconductivity is controlled 
by the phase-coherence). In other words, the optimal superconductivity is achieved in 
a regime where the physics is already that of  strong-coupling, with pair formation
occurring at a temperature higher than $T_c$.
We notice that the optimal interaction value is noticeably larger than the  attraction needed to form 
a bound state in the low-density limit  $U_b \sim D$\cite{randeria}.

\begin{figure}[htb]
\begin{center}
\includegraphics[width=8cm]{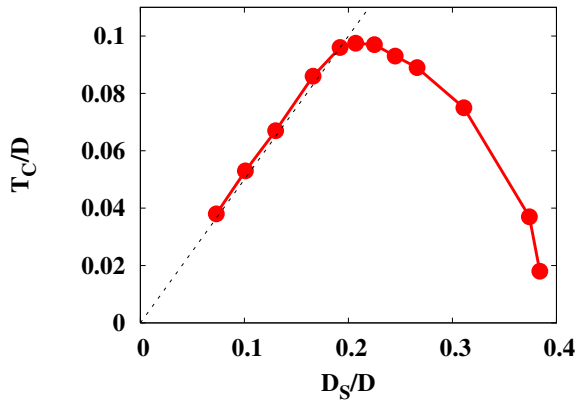}
\caption{\label{fig:uem_dmft} 
(Color online) Uemura plot in the attractive Hubbard model. The direct proportionality 
between $D_S$ and $T_c$ in the strong-coupling limit is evident on 
the left side of the figure, corresponding to $U > 2.4 D$.
} 
\end{center}
\end{figure}

Further insight on the influence of the crossing of the energy scales on
$T_c$ can be highlighted by plotting the critical temperature
versus $D_S$. In this way a sort of ``Uemura-plot'' relation, like 
that characterizing the physics of the underdoped cuprates\cite{uemura},
 is found in the  attractive Hubbard model for  $U > 2.4 D$, where $T_c$ is a decreasing function of $U$ 
(see Fig. \ref{fig:uem_dmft}).
This result supports the identification of the superconductivity with a condensation 
of preformed bosons  already at a moderate values of the  interaction.

On the other hand, the attractive Hubbard model, at least in our DMFT framework,
 cannot reproduce the deviations from the Uemura plot recently observed in the 
cuprates\cite{uem2002,panago}. 
More specifically  an evident ``re-entrance''  of the curve  $T_c(D_S)$  has been 
observed in many overdoped compounds,  and its shape appears to be strongly  material dependent. 
In our calculation at small $U$, where $D_S \simeq -\langle E_{kin} \rangle$, 
the superfluid stiffness is  a monotonically decreasing function of the interaction, and
does not show the experimentally observed re-entrance.

In order to push a little bit further the comparison with the most recent experimental 
data, it is useful to recast our DMFT results for the ``Uemura plot'' rescaling $T_c$ 
with the superconducting gap $\Phi_0$ in the single particle spectrum, 
as done by Tallon {\sl et al}\cite{panago}, in order to single out as much as 
possible the nonstandard behavior of $T_c$.
The experimental $T_c/\Phi_0$ is found
sub-linear in $D_S$ in the underdoped regime, while it is slightly super-linear at higher doping 
(with a  not universal tail in most overdoped compounds).
We notice that, beyond the BCS regime $\Phi_0$ can be different from the order parameter
$\Delta_0$ that we have introduced before. Here $\Phi_0$ is evaluated directly as the gap in the 
density of states. The difference between the two quantities varies up to 30\percent.

\begin{figure}[htb]
\begin{center}
\includegraphics[width=8cm]{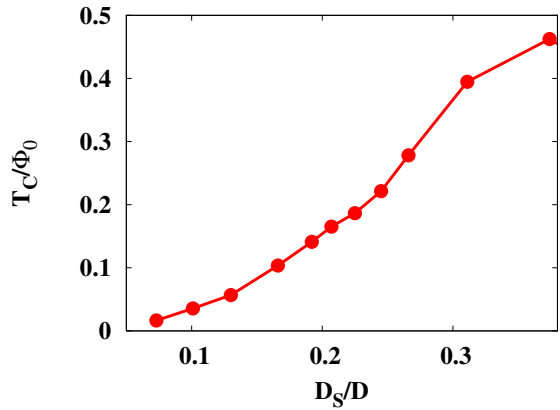}
\caption{\label{fig:panag_dmft} (Color online)  Rescaled ``Uemura-plot" (following Ref.  \onlinecite{panago})  
for  the attractive Hubbard model. 
We have plotted the  $T_c$ divided by the gap $\Phi_0$ as a function of  the 
superfluid density. It is evident  the sub-linear behavior of $T_c/\Phi_0$ in the intermediate-to-strong 
coupling regime, which mirror rather well the data of Ref. \onlinecite{panago} for several 
underdoped or  optimally doped  cuprates.}\end{center}
\end{figure}

Our results are summarized in Fig. \ref{fig:panag_dmft}, where one can note the presence of 
a wide  region of a sub-linear behavior of $T_c/\Phi_0$  on the left-side of the figure, which corresponds
to the intermediate-to-strong coupling regime in fair agreement with Ref. \onlinecite{panago}.
Also  a partial  of a super-linear behavior appears on the right-hand side of our plot.
The observed behavior of $T_c/\Phi_0$ is therefore captured by our simple BCS-BE crossover, 
without invoking more involved explanations\cite{panago}.

The evolution of energy scales we have discussed in this section
is reflected in an increased role of phase fluctuations as the
coupling grows. As we have mentioned in the introduction, those fluctuations  may have in principle  
relevant effects on the superconducting phase, in which finite dimensionality effects beyond DMFT can 
be important, possibly destroying the superconducting ordering. In DMFT, indeed,
phase fluctuations at large ${\bf q}$ are properly taken into account, 
while the small ${\bf q}$ collective modes are neglected.
In Ref. \onlinecite{kopec} it is argued that
superconductivity is destroyed by phase fluctuations for $U$ larger than a
critical value due to the small charge compressibility $\kappa = 
\partial n/ \partial \mu \propto 1/U$
derived within a phase-only effective theory based on the atomic limit, .
Since $\kappa$ measures the inertia of the system against the dynamic  
phase fluctuations, a small value would permit large zero-point fluctuations, 
eventually destroying phase coherence.
On the other hand, an alternative derivation of a phase-only action within 
one-loop expansion gives
instead $\kappa \sim 2U/D^2$, therefore increasing with $U$\cite{lungo}.

\begin{figure}[t!hb]
\begin{center}
\includegraphics[width=8cm]{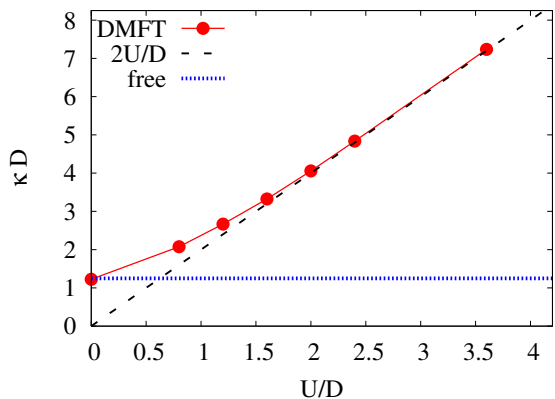}
\caption{\label{fig:chi_dmft} (Color online) Charge compressibility in the
 superconducting phase  at $T << T_c$. The
DMFT data are compared respectively with the strong-coupling behavior
from the phase-only action of Ref. \onlinecite{lungo}
and with the noninteracting value.}
\end{center} 
\end{figure}

Even if DMFT neglects finite-dimensionality effects and
cannot therefore treat the
physics of the Goldstone modes associated with the phase-fluctuations, 
we can use this approach to evaluate the coefficients appearing in
 the phase-only effective theories. The lack of bias towards weak or strong
coupling of DMFT should allow us to discriminate between the two derivations.
The results shown in  Fig. \ref{fig:chi_dmft}, clearly display that
$\kappa$ grows linearly with $U$ at strong coupling in agreement with  Ref. \onlinecite{lungo}:
the cost of dynamic phase fluctuation at ${\bf q}\simeq 0$ tends to increase in the 
large $U$ limit, leaving the superconducting phase stable, in contradiction with  Ref. \onlinecite{kopec}. 
A more complete description of the phase-fluctuations would require the 
calculation of all the phase-only theory coefficient, including the anharmonic ones.
This issue is beyond the aim of the present work.

\section{Energetic Balance of the  superconducting phase and 
optical sum rule}  

The characterization of the superconducting phase can 
be made more concrete by studying the temperature 
behavior of the kinetic ($E_{kin}$)  and the potential ($E_{pot}$) energies.
 From one side this analysis allows to evaluate
separately  all the energetic  contributions which characterize the 
mechanism for the stabilization of the superconducting long range order
 across the  BCS-BE crossover.
On the other hand, since in a lattice system the frequency integral 
of the optical conductivity  $\sigma(\omega)$ can be related 
to the kinetic energy of the carriers\cite{Vdm}, our
 calculations allow for a direct comparison
with the optical measurements on the cuprates.

\subsection{Kinetic and Potential Energy behavior}

The calculation of both potential and kinetic energies is quite 
straightforward in DMFT, since  it only involves local quantities. 
This is evident for the potential
energy, because by  definition it is proportional to the local density 
of double occupancies 
$n_d = \langle \sum_i n_{i\uparrow}n_{i\downarrow}\rangle$ 
\begin{equation}
E_{pot}= -U n_d. 
\end{equation}

As far as $E_{kin}$ is concerned, exploiting the simplified form 
of the self consistency equation for  the Bethe lattice 
we can derive the following expression
\begin{eqnarray}
E_{kin} & = & t^2  \, T \sum_{i \omega_n} [G(i \omega_n) \,
G_(i \omega_n) + G^{*}(i \omega_n) \,
G^{*}(i \omega_n)\nonumber \\  
\label{ekin_loc_expl}
& - & 2 F(i \omega_n) \,F(i \omega_n)],
\end{eqnarray}
which shows that also  the kinetic energy can be expressed only in terms of the local 
(normal and anomalous) Green function.

\begin{figure}[t!]
\includegraphics[width=82mm,height=77mm]{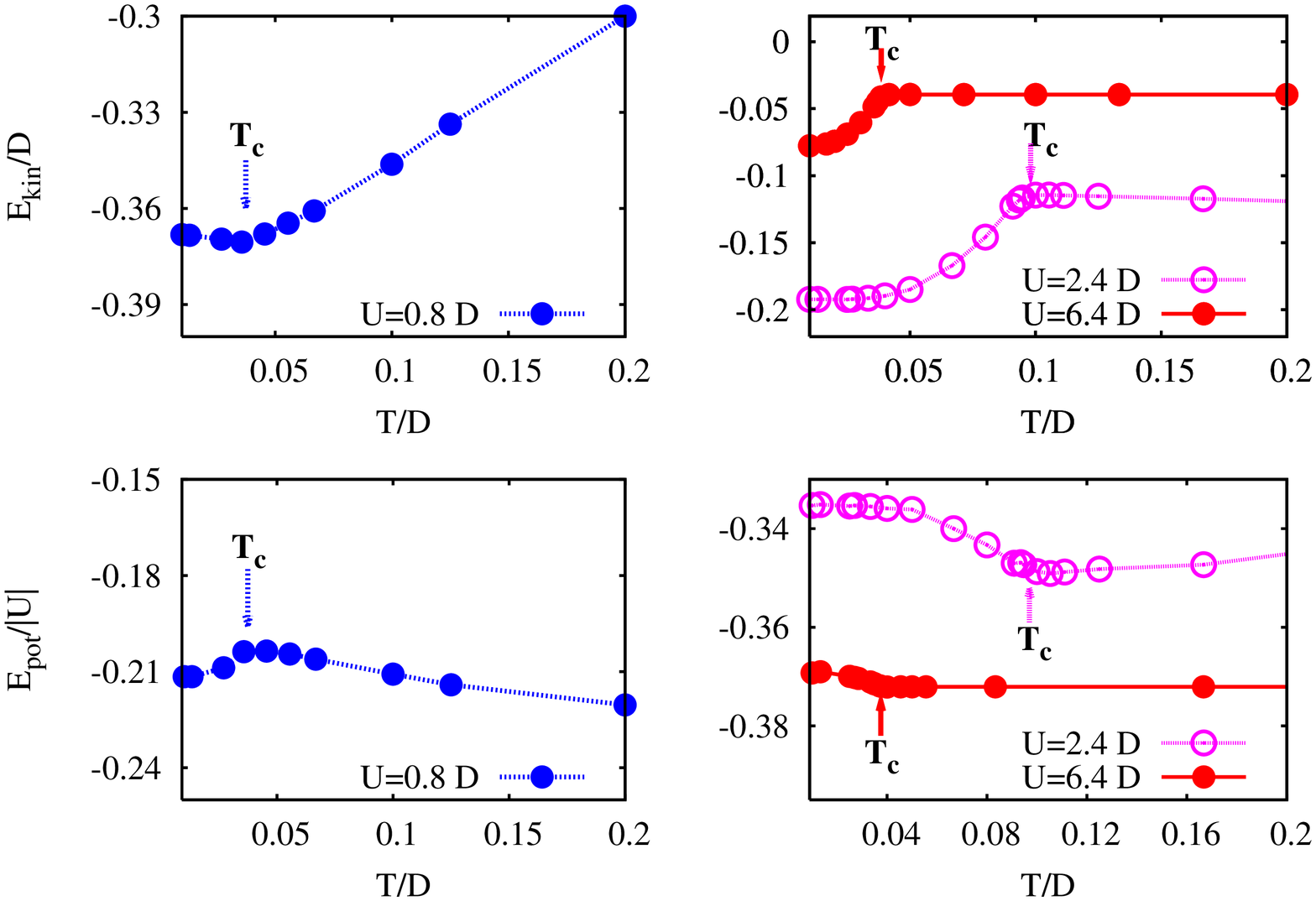}
\caption{\label{fig:ekin_dmft}
(Color online)  Low-temperature  behavior of the kinetic (upper panels) and the potential
(lower panels) energies. The critical temperature is 
marked by arrows. Notice that the kinetic energy is normalized by the half-bandwidth, and
the potential energy by the attraction strength $U$} 
\end{figure}

In Fig. \ref{fig:ekin_dmft}  we report
  the kinetic energy and potential energies for 
$U=0.8 D, 2.4D$ and $3.6D$, chosen as representative of the BCS, the intermediate
and the BE regimes, as a function of the temperature. 
The onset of the superconductivity is always marked   
 by an abrupt change in the concavity of $E_{kin}(T)$ e $E_{pot}(T)$. 
The energetic balance of superconductivity displays instead clear 
differences between the various  regimes.
At $U= 0.8 D$ the onset of superconductivity is accompanied by a slight
loss in the kinetic energy ($\Delta E_{kin}= E_{kin}(0)-E_{kin}(T_c)= +
 0.003 D$), and potential energy gain ($\Delta E_{pot}= 
E_{pot}(T=0)-E_{pot}(T_c) = -0.008 D$),  as in standard BCS theory.
On the other hand, 
both at $U= 2.4 D$ and $U = 6.4 D$ the superconducting 
phase is characterized by a lower value of $E_{kin}$ and 
 a loss of potential energy is observed below $T_c$. 

Such changes in the energetic balance at the
superconducting transition clearly highlight the different
mechanisms stabilizing superconductivity in the two regimes.
In the BCS limit superconductivity coincides with pair formation, which 
 determines a gain in potential energy and a consequent loss in kinetic
energy. In the opposite BE regime, the electrons are paired  at a high 
temperature of order $U$, but a true long-range superconducting order can 
take place only when phase coherence establishes between pairs.
Therefore $T_c$ is associated to a gain in kinetic energy, while a small
fraction of the potential energy gained with pair formation is lost at $T_c$.

\begin{figure}[b!]
\includegraphics[width=80mm,height=57mm]{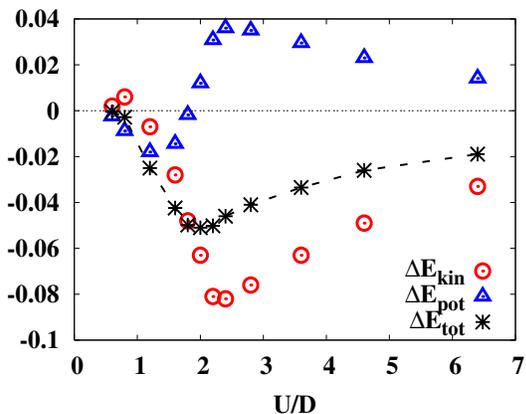}
\caption{\label{fig:ecvsU}
(Color online) 
 Kinetic, potential and total energy variation in the superconducting region 
for different values of the pairing interaction $U$} 
\end{figure}

More interestingly, 
our calculations  indicate that already at intermediate $U$ 
 the  ``strong-coupling'' mechanism is taking place, and superconductivity
is associated with a consistent  kinetic-energy gain, such as the one observed in 
the optimally doped cuprates (see next sub-section). 
The kinetic energy gain turns out to be maximum at intermediate $U/D$, where
the highest $T_c$ is achieved. 
The overall picture is drawn in Fig. \ref{fig:ecvsU} where we have reported 
the variation of the kinetic $\Delta E_{kin}$ , the potential $\Delta E_{pot}$ and the total 
energy $\Delta E_{tot}$ between $T=T_c$ and $T=0$ for a number of interaction values.
From these data one can easily see that {\it (i)} the typical BCS feature of an increasing
$E_{kin}$ below $T_c$ is rapidly lost (approximatively for $U \sim D \sim U_b$), 
in agreement with the discussion of the previous section; {\it (ii)} an  intermediate 
region exists,  in which the onset of the SC phase is associated with a gain of both 
potential and kinetic energy; {\it (iii)} 
the BE regime is effectively reached for $U$ larger than $1.8 D$
(hence before reaching the  maximal critical temperature), where superconductivity
is stabilized by kinetic energy.

It is also worth noting in Fig. \ref{fig:ecvsU} that the variation of the total energy $\Delta E_{tot}$
has approximatively the same dome-shape  behavior of the $T_c$ dependence on $U$. It is tempting then, to 
consider this value as a measure of the condensation energy $E_c=E_{tot}^N(T=0)-E_{tot}^S(T=0)$
 of the superconducting phase. This identification works well only if
one can assume a small variation of $E_{tot}^N(T)$ between $T=T_c$ and $T=0$, an assumption which holds
  both at small (due to the small value of $T_c$) and at intermediate-large $U$ (where both  
$E_{kin}$ and $E_{pot}$ are almost constant in temperature above $T_c$). However this is not the case
for $U \sim 1 \div 2$ D, where a strong temperature dependence of $E_{kin}(T)$ (and of $E_{pot}(T)$) in the 
normal phase is  found, because of the presence of a strongly renormalized quasiparticle excitation at
the Fermi  level\cite{paolo,lavoroopt}. Within DMFT the situation is even more involved 
for $ U_{c1}\sim  2.2 D < U < 2.9 D \sim U_{c2}$
where a coexistence between metallic and insulating solution is found and the temperature dependence 
of $E_{kin}$ and $E_{pot}$  in the normal phase below $T_c$ is subject to extremely abrupt 
changes\cite{paolo,prl}. 
However this is a peculiar feature of the DMFT treatment of the  Hubbard model which 
could not be so relevant 
when comparing our result with the experimental findings, also because these features of the 
Hubbard physics occur at temperatures  much smaller than $T_c$.

\subsection{Comparison with the optical measurements}

The integral over all the frequencies of the real part of the
optical conductivity in a lattice model is 
related to the second order derivative of the free particle dispersion\cite{hirsh,Vdm}
according to the following equation, 
\begin{eqnarray}  
\label{sum_expl1}
\int_{0}^{+\infty} \s_{xx}(\omega) \, d\omega &  = & \frac{\pi e^2}{2 N V}
\sum_{{\bf k},\s}  \frac{\partial^2 \e_{{\bf k}}}{\partial k_x^2} 
n_{{\bf k},\s} \\ \nonumber
 & = &  - \frac{\pi e^2 a^2} {2 d N V}
\sum_{{\bf k},\s} \e_{{\bf k}}  n_{{\bf k},\s}= - \frac{\pi e^2 a^2}{2 d V}  
\langle E_{kin} \rangle         
\label{sum_expl}
\end{eqnarray}
and the effect of the interactions is hidden in the values of the momentum distribution 
function $n_{{\bf k},\s}$.
Here $\langle E_{kin} \rangle=\frac{1}{N}\sum_{{\bf k},\s} 
\e_{\bf k} n_{{\bf k},\s}$ , $a$ the lattice spacing, $d$ the dimension of the system considered, 
and $V$  the unit-cell volume. The second equality, which holds for
nearest-neighbor hopping only, implies that the kinetic energy data in Figs. 
\ref{fig:ekin_dmft} and \ref{fig:ecvsU}  
can be compared directly with the the optical spectral weight 
of the conduction band in the cuprates.

The experimental results for  the low-energy behavior of the spectral weight behavior in
the HTS have been the object of a recent intense debate, and consensus has 
been reached about a few
general points. In  all the cuprates 
a sensible enhancement of  low-energy spectral weight $W(T) = $
$ \int_0^{\Omega_c} $ $\, d\omega \s(\omega,T)$ ($\Omega_C$ being a frequency cut-off of the
order of 1 eV, which is the plasma frequency for those materials),  is observed when 
lowering the temperature\cite{Vdm,Bont,homes,michele}. 
In the superconducting phase, underdoped (UD) and 
the optimally (Opt) 
doped samples behave differently from the overdoped (OD) ones. 
More precisely, in the UD-Opt compounds below $T_c$ an upward bump is observed 
with respect to the normal phase behavior, in contrast with the 
decreasing of $W(\Omega_c,T)$ 
observed in OD samples\cite{bont04}.

The gain of kinetic energy per $Cu$ atom 
$\Delta E_{kin} = E_{kin}(T_c) - E_{kin}(0)$ in  UD-Opt BSCCO can be estimated
using (\ref{sum_expl1}) as  $\Delta E_{kin} = - (0.5 \div 1)$ meV
\cite{Vdm,Bont,bont04}.       
Quite noticeably,  this value of 
$\Delta E_{kin}$ is much higher than the condensation energy for the same
material\cite{hirsh}, estimated as $E_c \sim 0.1$ meV  on the
 basis of the specific heat measurements\cite{specheat}.

According to our DMFT results, the bump of $W(\Omega_C,T)$ observed
 below $T_c$, can be qualitatively  understood in the
framework of the BCS-BE crossover. More specifically 
the hypothesis of an intermediate-to strong coupling description of the
superconductivity in the UD-Opt region 
appears to fit well with the observed enhancement of the 
low-frequency spectral  weight below $T_c$, whereas the disappearance of 
the upward bump in the OD cuprates is perfectly compatible with 
a  weak-coupling superconductivity.
It is interesting therefore to check to what extent is possible to push
forward such an analogy with the properties of the real systems. 

We can try to establish a  comparison with the experimental data 
by choosing the value of  the semi-bandwidth $D$ of the 
attractive Hubbard model in order to reproduce the maximum value $T_c^{max}$ 
(a $T_c^{max}$ of $90-100$ K is obtained with $D \sim 1000 K \sim 100 meV$).
The theoretical values for $\vert\Delta E_{kin}\vert$ and $\vert\Delta E_{tot}\vert$ 
(that we take as an estimate of $E_c$) are respectively $4 \div 8$ meV and $2 \div 4$ meV, larger by 
less  than one order of magnitude than the experimental values in  the underdoped HTSC.
It is anyway remarkable that the DMFT of the attractive Hubbard model correctly predicts the 
qualitative trend  of a condensation energy which is only a fraction of the kinetic energy gain.

As soon as the temperature exceeds $T_c$ and we enter the normal phase,
our calculation ceases to properly describe the experimental results.
The  $T^2$ behavior of $E_{kin}$ is in fact found only in the BCS region,
as shown by the $U=0.8 D$ data of Fig. \ref{fig:ekin_dmft}, and it completely disappears in 
the intermediate-to-strong coupling regime:
no relevant variation of $E_{kin}(T)$ is found at $U=2.4$, $6.4 D$ 
in a wide temperature range above $T_c$. 
The inadequacy of the present approach to describe the experimental findings 
in the normal phase has different origins. 
A first reason is the freezing of spatial fluctuations characteristic
of single-site DMFT. It is indeed reasonable to expect a role of short-range
fluctuations at least at strong coupling.
In this limit the attractive Hubbard model
can be mapped in an effective ``pseudospin'' hamiltonian\cite{micnasrev}, 
and the kinetic energy is given basically by  
nearest-neighbor correlations between the pseudospin operators, i.e.,
the local pairs and the empty sites. The temperature dependence of 
these correlations is poorly described by DMFT at large $U/D$. 
The relevance of this effect can tested using cluster extensions of 
DMFT\cite{jarrel,gabi}, where the short-range spatial correlations
are included.

There is however a more basic reason for the inadequacy of the attractive
Hubbard model to capture the temperature behavior of the normal phase
of the cuprates, which is the inability to properly describe the 
approach to the Mott insulator as the doping is reduced.
The underdoped region is indeed almost universally believed to be dominated
by the physics of a doped Mott insulator, with strongly renormalized
quasiparticles.
On the other hand, the strong-coupling region of our BCS-BE framework is certainly a 
``correlated'' regime, with renormalized quasiparticles, but the
interaction is simply the attractive one, which does not lead to the Mott insulating
state.
Indeed in a recent work\cite{lavoroopt} it has been  explicitly shown that 
the  physics of a doped  Mott insulating system, with its 
associated band-narrowing, may represent  the natural explanation for
 the strong temperature dependence of the spectral observed in the normal
phase of the cuprates.

\section{Conclusions}
 
In this paper we have performed a detailed investigation of the
physics of  superconducting phase in the attractive Hubbard model
 both at zero and at  finite temperature. The approach used here, namely the 
Dynamical Mean Field theory, is  completely non-perturbative allowing
for  a treatment of the superconducting phase properties, which is not
tied in principle either to the weak-coupling (BCS) or to the 
strong-coupling (BE) regime.

In particular we have investigated the behavior of the
energy scales relevant for the superconductivity as a function of the pairing 
interaction $U$. We have found that the evolution
of the superfluid density $D_S$ and the superconducting gap $\Delta_0$ 
displays a clear crossing in the intermediate coupling regime, 
 which is reminiscent of what is found experimentally in the phase-diagram
of HTSC for different doping levels.
In the weak-coupling region the BCS picture
works well and the superconductivity is controlled by the binding energy of the
Cooper pairs $\Delta_0$; on the contrary when the pairing interaction is
 high the superconductivity becomes essentially a phenomenon of
 superfluidity of preformed local pairs. 
 This is clearly witnessed by the direct proportionality
found in this regime between $T_c$ and $D_S$, and showed
 in a sort of ``Uemura-plot'' for 
the attractive Hubbard model. We also consider explicitly 
the problem of superconductivity suppression due to phase fluctuations in the hydrodynamic
regime, which  is usually neglected in a DMFT framework.
Our DMFT calculation of the compressibility
 indicates that, contrary to the claim of Ref. \cite{kopec}, the
phase-fluctuation effects do not destroy superconductivity, 
even for extremely large values of $U$.

The analysis of the superconducting 
phase properties in the attractive Hubbard model has been then further 
enriched by
considering the energetic stabilization of the superconductivity and 
the f sum-rule of the model. 
Our results clearly show a remarkable difference in the 
 energetic balance responsible for the stabilization of superconducting 
order when moving along the BCS-BE crossover. 
While in the weak-coupling regime the onset of
 superconductivity is associated to a gain of potential energy and a slight 
loss in kinetic energy as it is expected in a BCS picture, already in the
intermediate coupling at $U \simeq D$ and specifically  in proximity of the maximum of
 the $T_c$ dome, the situation is completely reversed.
 Here, and for higher values of $U$, the stabilization 
of the superconductivity is due to a marked reduction of $E_{kin}$, while
a smaller (but sizable) loss of potential energy is observed.
Therefore  the onset of superconductivity is clearly distinct
 from Cooper pair formation already for  moderate values of the pairing interaction. 

The energetic balance at the superconducting transition finds 
a direct experimental counterpart in the optical measurements, since
the integral of the optical conductivity is proportional to the kinetic energy itself. 
Our results can be rephrased in this terms, so that in weak-coupling a slight 
 reduction of the optical sumrule is found in our calculation at 
 the superconducting transition,
 while a relevant enhancement of the sumrule is observed for $T < T_c$ in
the whole intermediate-to-strong coupling region. 
This scenario is actually realized in optical measurements in the HTSC cuprates
 as a function of doping, with overdoped samples behaving more or less as standard BCS superconductors,
and underdoped one, characterized by a gain in kinetic energy\cite{Vdm,bont04,homes}.

A more quantitative comparison with the experiments reveals that the 
variations of both kinetic energy and condensation energy 
between $T=0$ and $T_c$ are reasonably well reproduced by our model, 
which also correctly predicts that a sizable fraction of the kinetic
energy gain is canceled by the potential energy loss, leading to a condensation
energy significantly smaller than the kinetic energy gain, as 
 it indeed happens\cite{hirsh}  in the HTSC.
It must be noticed that our calculation tends to overestimate both kinetic
energy gain and condensation energy.
However, the main limitation of the attractive Hubbard model in this context is the
inability to capture the  $T^2$ enhancement of the spectral
weight in the normal phase above $T_c$. The main reason for this 
disagreement is likely the complete neglect of the strong repulsive
interactions characterizing the underdoped cuprates close to the Mott state.

\begin{figure}[hbt]
\includegraphics[width=8cm,angle=0]{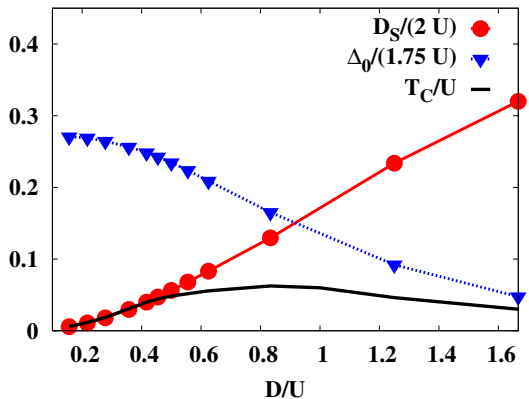}
\caption{\label{fig:cross_vsD} 
(Color online) Same quantities of Fig. 2 plotted as a function of $D/U$. Both $D_S$ and $\Delta_0$ 
are normalized by $U$. Now the crossing between the 
superconducting gap $\Delta_0$ 
and the superfluid stiffness at $D \sim U$, 
occurs in the proximity of the maximum value of $T_c$.} 
\end{figure}

An important outcome of our analysis is that the optimal critical 
temperature is
obtained for an interaction strength well inside the BE region.
This finding is in good agreement with previous calculations of the pseudogap 
temperature $T^*$ extracted from the spin susceptibility and specific heat, where
the $T^*$ is significantly larger than $T_c$ at optimal interaction\cite{paolo},
but it is in contrast with experiments in Bismuth and Yttrium based cuprates,
where $T^*$ tends to vanish close to optimal doping.
It is interesting to notice that this discrepancy seems to disappear if we 
fix the attraction strength $U$ and follow the BCS-BE crossover by varying the 
half-bandwidth $D$, and measuring energies in units of $U$. 
Such a rescaling is meant to roughly describe a situation in which 
the crossover from BCS to BE is due to a shrinking of the coherent band 
due to strong repulsive correlation effects which are stronger and stronger
as the Mott insulator is approached, while the attraction, which we might 
think to arise from antiferromagnetic superexchange for the sake of definiteness, is basically unrenormalized in the same process.

As shown in Fig. \ref{fig:cross_vsD}, the maximum 
critical temperature now occurs for $U\simeq D$, at the boundary of the BCS
region, where the pseudogap is small.
However this different perspective, in which the optimal $T_c$ moves closer to the BCS
region is in contradiction with the experimental evidence of a kinetic-energy
driven superconductivity around optimal doping, which is qualitatively obtained in the 
BE regime.

As a  summary of our results, we can draw some final 
considerations on the the connection of the BCS-BE crossover scenario  
and the physics of the HTSC:
several qualitative feature of the HTSC properties (as the crossing of the 
hierarchy of $D_S$ and $\D_0$, the Uemura plot ,\ldots) 
can be actually captured within a purely attractive 
description. However, 
as a general trend, it provides
too high estimation of the values of many thermodynamic quantities as the 
kinetic energy variation below $T_c$ and the condensation energy. 
At the same time the purely
attractive description fails in reproducing the magnitude of the temperature 
dependence of the optical spectral weight above $T_c$, 
calling for the inclusion of  strong correlation
effect and suggesting that a closer agreements with the physics of the 
cuprates may be obtained by applying the attractive Hubbard description,
or more generally the BCS-BE crossover picture,
only to the narrow quasiparticle excitations 
characteristic of a doped Mott insulator.
Models in which explicit attractive and repulsive interaction are present\cite{capone1,capone2}
are natural candidates to explore this scenario.

After completion of this work, we became aware of recent Cellular 
DMFT\cite{gabi}
 results by Kyung, Georges and Tremblay (cond-mat/0508645), 
where it is proposed that 
the inclusion of short range correlations beyond single-site DMFT
makes the evolution on the normal state continuous, and the zero-temperature 
energetic balance
of the superconducting transition is studied.

\section{acknowledgements}  
We acknowledge useful discussions with A. Georges, 
M. Ortolani and  L. Benfatto, as well as financial support
by Miur Cofin 2003 and CNR-INFM..


\begin{thebibliography}{99}
\bibitem{notadiam} This relation requires  two conditions: (i) the kinetic
term must be given by a next-neighbour hopping and (ii) in  
the electronic interaction  terms no coupling with e.m. potential 
must be present.
\bibitem{notadim} This is dictated by the fact that in 
 the $d \rightarrow \infty$ limit, the contribution to $D_{dia}$ and 
$\chi_{jj}$ of a single direction vanishes as 
$1/d$, due to the scaling of the hopping term $t \rightarrow t/\sqrt{d}$.
\bibitem{notadefds} We have expressed  the superfluid 
stiffness as an energy in order to make more 
transparent the comparison with  the order parameter $\Delta_0$.
\end{thebibliography}
\end{document}